\documentclass[12pt]{article}
\usepackage{amsmath,amsfonts,xypic, epsfig}
\usepackage{amssymb}
\usepackage{amscd}
\advance \textheight by \topskip
\makeatletter

\@addtoreset{equation}{section}
 \def\theequation{\thesection.\arabic{equation}}
\makeatother

\newtheorem{theorem}{Theorem}[section]

\newtheorem{definition}[theorem]{Definition}

\def\NN{{\mathbb N}}

\def\RR{{\mathbb R}}
\def\NN{{\mathbb N}}

\newcommand{\beqa}{\begin{eqnarray}}
\newcommand{\eeqa}{\end{eqnarray}}
\newcommand{\noi}{\noindent}

\def\>{\rangle}
\def\<{\langle}

\begin{document}

\title{
{\bf Parametric representation of a translation-invariant renormalizable noncommutative model
}  }
\author{ 
{\sf Adrian Tanasa}\thanks{e-mail:
adrian.tanasa@ens-lyon.org}$\,\,$$^{a,b}$
\\
{\small ${}^{a}${\it Centre de Physique Th\'eorique, CNRS UMR 7644,}} \\
{\small {\it Ecole Polytechnique, 91128 Palaiseau, France}}  \\ 
{\small ${}^{b}${\it Departamentul de Fizic\u a Teoretic\u a,}}\\
{\small {\it Institutul de Fizic\u a \c si Inginerie Nuclear\u a ``Horia Hulubei'',}} \\
{\small {\it P. O. Box MG-6, 077125 Bucure\c sti-M\u agurele, Rom\^ania}}  \\ 
}
\maketitle
\vskip-1.5cm

\vspace{2truecm}

\begin{abstract}
\noindent
We construct here the parametric representation of a translation-invariant renormalizable scalar model on the noncommutative Moyal space of even dimension $D$. This representation of the Feynman amplitudes is based on some integral form of the noncommutative propagator. All types of graphs (planar and non-planar) are analyzed. The r\^ole played by noncommutativity is explicitly shown. This parametric representation established allows to calculate the power counting of the model.
Furthermore, the space dimension $D$ is just a parameter in the formulas obtained. This paves the road for the dimensional regularization of this noncommutative model.
\end{abstract}

\bigskip




\newpage

\section{Introduction and motivation}
\renewcommand{\theequation}{\thesection.\arabic{equation}}   
\setcounter{equation}{0}

Noncommutative geometry (see \cite{book-connes}) is an appealing framework for the quantization of gravity. One can thus address the question of weather or not space-time becomes noncommutative at energy scales approaching the Planck scale \cite{gedanken}. Moreover,  relations of noncommutativity and string theory \cite{string1, string2} or loop quantum gravity \cite{loop} have been established.

Nevertheless, noncommutative quantum field theory (NCQFT) - for a general review see \cite{DN}- is known to suffer, at least on the Moyal space, from a new type of divergence, the UV/IR mixing \cite{melange}. This divergence is responsible for the non-renormalizability of NCQFT. 
The Grosse-Wulkenhaar scalar model  \cite{GW1, GW2}
is a first solution to this problem, restoring the perturbative renormalization \cite{GW1, GW2, Orsay, GW4, dimreg, hopf}. Moreover, due to a vanishing $\beta$ function at all orders in perturbative theory \cite{beta1, beta23, beta}, a constructive version of this model is within reach \cite{constructiva}. 
Several other field theoretical results have been established for this  model and other related noncommutative models \cite{param1, param, mellin, param2, goldstone} - for some general reviews see \cite{raimar, vince}. Let us emphasize that among these results one has the parametric representation \cite{param1, param, param2}; based on this, one further has the Mellin representation of the Feynman amplitudes as well as the implementation of the dimensional regularization scheme \cite{dimreg}.

Nevertheless, the Grosse-Wulkenhaar model explicitly breaks the translation-invariance of field theory. 
Furthermore, it does not seem easy to generalize this method to gauge theories: one is lead to theories with non-trivial vacua \cite{gauge}, in which renormalizability is unclear up to now.

A different scalar model on the Moyal space was proposed in \cite{noi} and proved renormalizable at any order in perturbation theory. Moreover this model is manifestly translation-invariant. Its one-loop renormalization group flows were computed in \cite{beta-GMRT}. Let us also mention that the extension of this new method to gauge theories has been recently proposed \cite{gauge2}.

In this paper we establish the parametric representation of this latter model. An integral representation of the propagator is used, representation which splits the noncommutative propagator into a sum of the ``usual'' commutative propagator and some noncommutative correction. This splitting further leads to a splitting of the Feynman amplitude into $2^L$ terms, where $L$ is the number of internal lines of the graph. We investigate in this paper all types of graphs, planar or non-planar. This allows one to observe the r\^ole played by noncommutativity in each of these cases. The parametric representation formulas we establish allow one to confirm the power counting obtained in \cite{noi}, where a different method, the multi-scale analysis is used. Furthermore, let us also emphasize that the space dimension $D$ is just a parameter in the formulas obtained. (Recall that this was also the case in  commutative theory). This paves the road for the implementation of the dimensional regularization scheme. Moreover, the results of this paper also allow the definition of an appropriate Mellin representation of the model \cite{noi}, as it was done for the Grosse-Wulkenhaar model.

The paper is structured as follows. The next section recalls the parametric representation of the commutative $\Phi^4$ model. The third section recalls some notions of topology of ribbon Feynman graphs (the graphs used in NCQFT); we define here the notion of regular graphs. Furthermore we briefly recall the noncommutative model we study. The decomposition of the noncommutative propagator into a sum of the commutative propagator and some noncommutative correction is given. In the following section we analyze the planar regular graphs and we establish their parametric representation. The next two sections are devoted to the more involved cases of planar irregular and resp. non-planar graphs. In the last section we give some examples from all the categories of graphs above,  planar (regular or irregular) or non-planar.

\section{The $\Phi^4$ model on the commutative $\RR^4$; its parametric representation}
\renewcommand{\theequation}{\thesection.\arabic{equation}}   
\setcounter{equation}{0}

Let us give here the results of the parametric representation for commutative
quantum field theory (one can see for example 
\cite{carte} for further details).

To any internal line of a graph $G$ one associates a parameter $0<\alpha<\infty$ {\it via} the integral representation of the respective propagator
\beqa
\label{propa-com}
C(p,m)=\frac{1}{p^2+m^2}=\int_0^\infty d\alpha e^{-\alpha(p^2+m^2)}.
\eeqa

To any graph $G$ one associates the polynomials $U$ and $V$ defined as:

\begin{definition}
\label{polinoame}
Let the following polynomials depending on the set of parameters $\alpha_i$ ($i=1,\ldots,L$)
and on the set of external momenta $p$
\beqa
\label{s1}
U (\alpha)&=& \sum_{\cal T} \prod_{\ell \not \in {\cal T}} \alpha_\ell \ , \nonumber\\
V (\alpha, p)&= &\sum_{{\cal T}_2} \prod_{\ell \not \in {\cal T}_2} \alpha_\ell  (\sum_{i \in
  E({\cal T}_2)} p_i)^2 \ , 
\eeqa
where $\cal T$ is a (spanning) tree of the graph and  ${\cal T}_2$ is a
$2-$tree, ({\it i. e.} a tree minus one of its lines) which separates the graph in two connected components, one of them being  $E({\cal T}_2)$.
\end{definition}

The amplitude of a Feynman graph $G$ writes
\beqa \label{as} 
{\cal A}_G = \pi^{\frac {LD}{2}}\int_0^{\infty} 
\frac{e^{- V(\alpha,p)/U (\alpha) }}{U (\alpha)^{D/2}} 
\prod_{\ell=1}^L  ( e^{-m^2 \alpha_\ell} d\alpha_\ell )\ .
\eeqa

Finally, let us also emphasize that by a rescaling of the parameters $\alpha$ one can obtain the superficial degree of divergence $\omega$ of the model. 

\section{Feynman ribbon graphs. The noncommutative model}
\renewcommand{\theequation}{\thesection.\arabic{equation}}   
\setcounter{equation}{0}
\label{Feynman}

\subsection{Feynman ribbon graphs: planarity and non-planarity, rosettes}

\begin{figure}
\centerline{\epsfig{figure=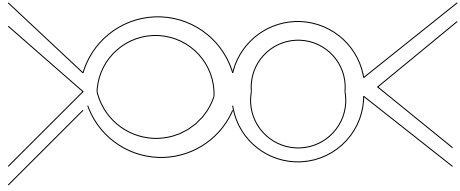,width=6cm} }
\caption{An example of a ribbon graph. This graph has $2$ loops and is planar regular.}
\label{ribbon}
\end{figure}

In this section we give some useful conventions and  definitions.
As already stated in the previous section, the Feynman graphs used in NCQFT are ribbon graphs.
Let us consider such a $\Phi^4$ graph with $n$ vertices, $L$ internal lines and $F$
faces (see for example the Feynman graph of Fig. \ref{ribbon}).
One has
\beqa
\label{genus}
2-2g=n-L+F,
\eeqa
\noi
where $g\in\NN$ is the {\it genus} of the manifold on which the respective ribbon graph is drawn.
If $g=0$ we call the respective graph a {\it planar graph}, if $g>0$ we talk about a {\it non-planar
  graph}. Furthermore, we call a planar graph to be a {\it planar regular
  graph} if it has no faces broken by external lines. We denote the number of faces broken by external lines by $B$.

\begin{figure}
\centerline{\epsfig{figure=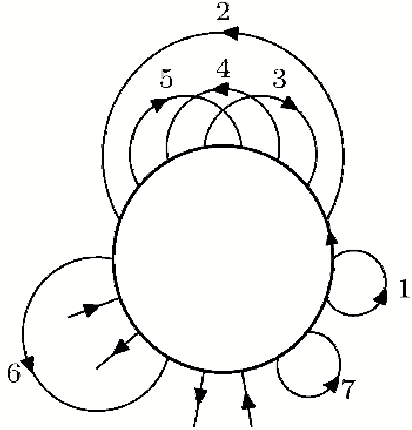,width=6cm} }
\caption{An example of a rosette. The crossings of lines $3$, $4$ and $5$ indicate that one deals with a non-planar graph. Furthermore, the face overarched by line $6$ is broken by two external legs.}\label{roz}
\end{figure}

In \cite{filk}, several contractions on such a Feynman 
graph were defined. The one we use in the sequel consists in reducing a tree line by gluing up
together two vertices into a bigger one.
Repeating this operation for the $n-1$ tree lines, one obtains a single final
vertex with all the loop lines hooked to it - a {\it rosette} (see Fig. \ref{roz}).

The number of faces or the genus of the graph do not change 
under this operation. Furthermore, the external legs will break the same faces on the rosette as they were breaking them on the initial Feynman graph. When one deals with a planar graph, there will be no crossing between the loop lines on the rosette. The example of Fig. \ref{roz} corresponds thus to a non-planar graph (one has crossings between the loop lines $3$, $4$ and $5$). 

In \cite{filk} there were further indicated the general oscillating factors appearing (for a theory on the Moyal space) in the Feynman integrand as a function of the type of the corresponding rosette (and thus of planarity and number of broken faces). We do not recall here all these results. We will however come back on this particular factors when requested in the proofs that will follow.

\subsection{The noncommutative $\Phi^4$ model}

We place ourselves on the $D-$dimensional Moyal space ($D$ even)
\beqa
\left[x_\mu,x_\nu\right]{}_\star=i\Theta_{\mu\nu},
\eeqa
where $[x_\mu,x_\nu]=x_\mu\star x_\nu - x_\nu\star x_\mu$ and $\star$ denotes the Moyal product. The noncommutativity matrix $\Theta$ is block-diagonal, each block writing as 
\beqa
\label{matrix}
\begin{pmatrix}
0 & \theta \\
-\theta & 0
\end{pmatrix}
\eeqa
In \cite{noi}, the following model was introduced
\beqa
\label{revolutie}
S_\theta [\phi]=\int d^D p (\frac 12 p_{\mu} \phi  p^\mu \phi  +\frac
12 m^2  \phi  \phi   
+ \frac 12 a  \frac{1}{\theta^2 p^2} \phi  \phi  
+ \frac{\lambda }{4!} V_\theta ),
\eeqa
with  $a $  some dimensionless parameter. By $V_\theta$ we understand the corresponding potential $\frac{\lambda}{4!}\phi(x)^{*4}$ in momentum space.
The propagator is
\beqa
\label{propa}
C(p,m,\theta)=\frac{1}{p^2+m^2+\frac{a}{\theta^2 p^2}} \, ,
\eeqa
We further choose $ \frac 14 \theta^2 m^4 \ge a > 0$.

\medskip

In order to obtain the parametric representation associated to the model \eqref{revolutie}, one needs an integral representation of the propagator \eqref{propa}. 
We give this representation following \cite{beta-GMRT} 
\beqa
\label{propa2}
C(p,m,\theta)&=&\frac{1}{p^2+m^2}-\frac{1}{p^2+m^2}\frac{a}{\theta^2 p^2 (p^2+m^2)+a}\nonumber\\
&=&
\frac{1}{p^2+m^2}-\frac{1}{p^2+m^2}\frac{a}{\theta^2 (p^2 +m_1^2)(p^2+m^2_2)},
\eeqa
where $-m_1^2$ and $-m_2^2$ are the roots of the denominator of the second term in the RHS of the first line above (considered as a second order equation in $p^2$, namely
$\frac{-\theta^2 m^2\pm \sqrt{\theta^4 m^4 - 4 \theta^2 a}}{2\theta^2}<0$. 
The form \eqref{propa2} allows to write an integral representation of the propagator $C(p,m,\theta)$. Nevertheless, for the second term one needs a triple integration over some set of appropriate Schwinger parameters:
\beqa
\label{param}
C(p,m,\theta)&=&\int_0^\infty d\alpha e^{-\alpha (p^2+m^2)}\nonumber\\
&-& \frac{a}{\theta^2} \int_0^\infty \int_0^\infty d\alpha d\alpha^{(1)} d\alpha^{(2)} e^{-(\alpha+\alpha^{(1)}+\alpha^{(2)})p^2} e^{-\alpha m^2} e^{-\alpha^{(1)} m_1^2}e^{-\alpha^{(2)} m_2^2}.
\eeqa

One can further continue the decomposition of the last term in \eqref{propa2} in simple elements. Nevertheless, the total number of Schwinger parameters $\alpha$ requested is also four. In this paper we will work out the parametric representation using the form \eqref{param}.





\bigskip

We now proceed with the implementation of the parametric representation. We start with the case of planar regular graphs, then continue with the planar irregular graphs and finally end up with the non-planar ones.

The phase obtained from Moyal oscillations coupling the external momenta to themselves is not taken into consideration. This phase interferes only in a trivial way in the Gaussian integrations over internal momenta (it actually factorizes out).

\section{Planar regular graphs}
\renewcommand{\theequation}{\thesection.\arabic{equation}}   
\setcounter{equation}{0}

\begin{figure}
\centerline{\epsfig{figure=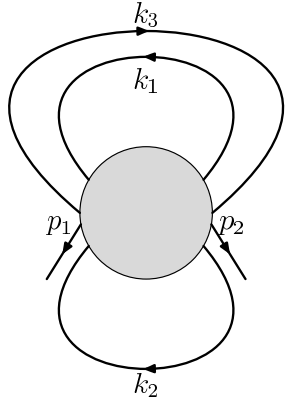,width=3.7cm} }
\caption{A planar regular rosette. One has no internal faces broken by external legs 
($B=1$) and no crossing between the loop lines ($g=0$).}\label{roz-plr}
\end{figure}

The parametric representation of a Feynman amplitude of some planar regular graph is given by the following Theorem:

\begin{theorem}
\label{th-plr}
Let $G$ a planar regular graph. Its Feynman amplitude writes
\beqa
\label{plr}
&&{\cal A}_G=K_G \left( \int \prod_{i=1}^L d \alpha_i \frac{1}{[U(\alpha)]^{\frac D2}} e^{\frac{-V(\alpha, p)}{U(\alpha)}} e^{-\sum_{i=1}^L \alpha_i m^2 }\right.\\
&& + (-\frac{a}{\theta^2})^{L-1}\sum_{j_1=1}^L \int d\alpha_{j_1} \prod_{i\ne j_1,\, i=1}^L d\alpha_i d\alpha_i^{(1)} d\alpha_i^{(2)} \frac{1}{[U(\alpha_i+\alpha_i^{(1)}+\alpha_i^{(2)}, \alpha_{j_1})]^\frac D2} e^{-\frac{V(\alpha_i+\alpha_i^{(1)}+\alpha_i^{(2)}, \alpha_{j_1},p)}{U(\alpha_i+\alpha_i^{(1)}+\alpha_i^{(2)}, \alpha_{j_1})}} \nonumber\\
&& \ \ \ \ \ \ \ \ \ \ \ \ \ e^{-\sum_{i=1}^L \alpha_i m^2 } e^{-\sum_{i\ne j_1,\, i=1}^L \alpha_i^{(1)} m_1^2 }e^{-\sum_{i\ne j_1,\, i=1}^L \alpha_i^{(2)} m_2^2 }\nonumber\\
&& + (-\frac{a}{\theta^2})^{L-2}\sum_{j_1<j_2,\, j_1,j_2=1}^L \int d\alpha_{j_1}d\alpha_{j_2} \prod_{i\ne j_1,j_2,\, i=1}^L d\alpha_i d\alpha_i^{(1)} d\alpha_i^{(2)} \frac{1}{[U(\alpha_i+\alpha_i^{(1)}+\alpha_i^{(2)}, \alpha_{j_1}, \alpha_{j_2})]^\frac D2} \nonumber\\
&& \ \ \ \ \ \ \ \ \ \ e^{-\frac{V(\alpha_i+\alpha_i^{(1)}+\alpha_i^{(2)}, \alpha_{j_1},\alpha_{j_2}, p)}{U(\alpha_i+\alpha_i^{(1)}+\alpha_i^{(2)}, \alpha_{j_1}, \alpha_{j_2})}}
e^{-\sum_{i=1}^L \alpha_i m^2 } e^{-\sum_{i\ne j_1j_2,\, i=1}^L \alpha_i^{(1)} m_1^2 }e^{-\sum_{i\ne j_1,j_2\, i=1}^L \alpha_i^{(2)} m_2^2 }
\nonumber\\
&& + \ldots +\nonumber\\
&&  + (-\frac{a}{\theta^2})^{L}\int  \prod_{i=1}^L d\alpha_i d\alpha_i^{(1)} d\alpha_i^{(2)} \frac{1}{[U(\alpha_i+\alpha_i^{(1)}+\alpha_i^{(2)})]^\frac D2} e^{-\frac{V(\alpha_i+\alpha_i^{(1)}+\alpha_i^{(2)},p)}{U(\alpha_i+\alpha_i^{(1)}+\alpha_i^{(2)})}}\nonumber\\
&& \left. \ \ \ \ \ \ \ \ \ \ \ \ \ \ \ \ \ \ \ \ \ \ \ \ \ \ \ 
e^{-\sum_{i=1}^L \alpha_i m^2 }e^{-\sum_{i=1}^L \alpha_i^{(1)} m_1^2 }e^{-\sum_{i=1}^L \alpha_i^{(2)} m_2^2 }
\right).\nonumber
\eeqa
\end{theorem}
{\it Proof:} Since we deal with a planar regular graph which has a rosette of the type of Fig. \ref{roz-plr}, one has no oscillator factor. 
The Feynman amplitude of $G$ writes
\beqa
\label{a-plr}
\int \prod_{i=1}^L d^D k_i C(k_i, m, \theta) \delta^D (\sum k,p), 
\eeqa
where the $\delta$ function expresses the momentum conservation at each vertex.
Inserting the integral representation \eqref{param} in \eqref{a-plr} one has
\beqa
&&\int \prod_{i=1}^L d^D k_i \delta^D (\sum k,p) \\
&&\left(  \int d \alpha_1 e^{-\alpha_1 k_1^2} e^{-\alpha_1 m^2} 
- \frac{a}{\theta^2}  \int d \alpha_1 d\alpha_1^{(1)}d\alpha_1^{(2)} e^{-(\alpha_1+\alpha_1^{(1)}+\alpha_1^{(2)}) k_1^2}e^{-\alpha_1 m^2} e^{-\alpha_1^{(1)} m_1^2}e^{-\alpha_1^{(2)} m_2^2} \right)\nonumber\\
&&\left(  \int d \alpha_2 e^{-\alpha_2 k_2^2} e^{-\alpha_2 m^2} 
- \frac{a}{\theta^2}  \int d \alpha_2 d\alpha_2^{(1)}d\alpha_2^{(2)} e^{-(\alpha_2+\alpha_2^{(1)}+\alpha_2^{(2)}) k_2^2}e^{-\alpha_2 m^2} e^{-\alpha_2^{(1)} m_1^2}e^{-\alpha_2^{(2)} m_2^2} \right)\nonumber\\
&&\ldots\nonumber\\
&&\left(  \int d \alpha_L e^{-\alpha_L k_L^2} e^{-\alpha_L m^2} 
- \frac{a}{\theta^2}  \int d \alpha_L d\alpha_L^{(1)}d\alpha_L^{(2)} e^{-(\alpha_L+\alpha_L^{(1)}+\alpha_L^{(2)}) k_L^2}e^{-\alpha_L m^2} e^{-\alpha_L^{(1)} m_1^2}e^{-\alpha_L^{(2)} m_2^2} \right).\nonumber
\eeqa
Expanding the product above leads to $2^L$ terms, listed in \eqref{plr} as a polynomial (of degree $L$) in $\frac{a}{\theta^2}$. The first of them corresponds to the usual parametric representation of the commutative $\Phi^4$ model, as expected. This was obtained by performing the Gaussian integration on the set of internal momenta $k_i$. The Gaussian integrations in the rest of the $2^L-1$ terms are performed analogously; one just needs to 
make the appropriate
replacements
$$\alpha_{i}\to \alpha_{i}+\alpha_{i}^{(1)}+\alpha_{i}^{(2)}$$
and integrate over the corresponding set of $\alpha$ parameters, 
as indicated in \eqref{plr}. (QED)

\medskip

For the planar regular graph the noncommutativity translates only in some corrections to the usual commutative parametric representation, correction coming form the form \eqref{propa} of the noncommutative propagator. This is a direct consequence of the fact that, in this case, one has not any oscillating factor mixing the internal and external momenta {\it via} the noncommutativity matrix $\Theta$ (see \cite{filk} for further details on this issue).



\section{Planar irregular graphs}
\label{sect-pli}
\renewcommand{\theequation}{\thesection.\arabic{equation}}   
\setcounter{equation}{0}

Let us first investigate the case of two broken faces, $B=2$ and then the more involved one of an arbitrary number, $B\ge 3$, of broken faces.

\subsection{Two broken faces}

\begin{figure}
\centerline{\epsfig{figure=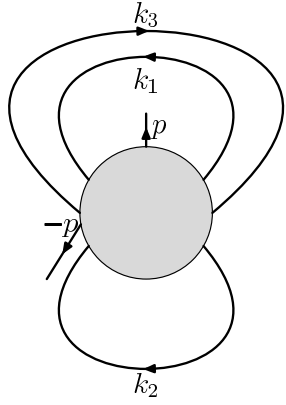,width=3.7cm} }
\caption{A two point, two broken faces rosette. One internal face is broken by the momentum $p$. This face is overarched by the momenta $k_1-k_3$.}\label{rozeta-2P-B2}
\end{figure}

\begin{theorem}
\label{th-plB2}
Let $G$ a planar irregular graph, with two broken faces, one external and one internal. 
Let $\sum p_k$ be the sum of the external momenta (with their corresponding signs) breaking the internal face and let $\sum \alpha_\ell$ the sum of the parameters corresponding to the tree lines and to the lines overarching the face broken by the momenta $p_k$ in the rosette. 
The Feynman amplitude writes
\beqa
\label{plB2}
&&{\cal A}_G=\pi^{\frac{LD}{2}} \left( \int \prod_{i=1}^L d \alpha_i \frac{1}{[U(\alpha)]^{\frac D2}} e^{\frac{-V(\alpha, p)}{U(\alpha)}} 
e^{-\frac{\theta^2 (\sum p_k)^2}{4\sum \alpha_\ell}}
e^{-\sum_{i=1}^L \alpha_i m^2 }\right.\\
&& + (-\frac{a}{\theta^2})^{L-1}\sum_{j_1=1}^L \int d\alpha_{j_1} \prod_{i\ne j_1,\, i=1}^L d\alpha_i d\alpha_i^{(1)} d\alpha_i^{(2)} \frac{1}{[U(\alpha_i+\alpha_i^{(1)}+\alpha_i^{(2)}, \alpha_{j_1})]^\frac D2} e^{-\frac{V(\alpha_i+\alpha_i^{(1)}+\alpha_i^{(2)}, \alpha_{j_1},p)}{U(\alpha_i+\alpha_i^{(1)}+\alpha_i^{(2)}, \alpha_{j_1})}} \nonumber\\
&& \ \ \ \ \ \ \ \ \ \ \ \ \ e^{-\frac{\theta^2 (\sum p_k)^2}{4\sum \alpha_\ell}}
e^{-\sum_{i=1}^L \alpha_i m^2 } e^{-\sum_{i\ne j_1,\, i=1}^L \alpha_i^{(1)} m_1^2 }e^{-\sum_{i\ne j_1,\, i=1}^L \alpha_i^{(2)} m_2^2 }\nonumber\\
&& + (-\frac{a}{\theta^2})^{L-2}\sum_{j_1<j_2,\, j_1,j_2=1}^L \int d\alpha_{j_1}d\alpha_{j_2} \prod_{i\ne j_1,j_2,\, i=1}^L d\alpha_i d\alpha_i^{(1)} d\alpha_i^{(2)} \frac{1}{[U(\alpha_i+\alpha_i^{(1)}+\alpha_i^{(2)}, \alpha_{j_1}, \alpha_{j_2})]^\frac D2} \nonumber\\
&& \ \ \ \ \ \ \ \ \ \ e^{-\frac{V(\alpha_i+\alpha_i^{(1)}+\alpha_i^{(2)}, \alpha_{j_1},\alpha_{j_2}, p)}{U(\alpha_i+\alpha_i^{(1)}+\alpha_i^{(2)}, \alpha_{j_1}, \alpha_{j_2})}}
e^{-\frac{\theta^2 (\sum p_k)^2}{4\sum \alpha_\ell}}
e^{-\sum_{i=1}^L \alpha_i m^2 } e^{-\sum_{i\ne j_1j_2,\, i=1}^L \alpha_i^{(1)} m_1^2 }e^{-\sum_{i\ne j_1,j_2\, i=1}^L \alpha_i^{(2)} m_2^2 }
\nonumber\\
&& + \ldots +\nonumber\\
&&  + (-\frac{a}{\theta^2})^{L}\int  \prod_{i=1}^L d\alpha_i d\alpha_i^{(1)} d\alpha_i^{(2)} \frac{1}{[U(\alpha_i+\alpha_i^{(1)}+\alpha_i^{(2)})]^\frac D2} e^{-\frac{V(\alpha_i+\alpha_i^{(1)}+\alpha_i^{(2)},p)}{U(\alpha_i+\alpha_i^{(1)}+\alpha_i^{(2)})}}
e^{-\frac{\theta^2 (\sum p_k)^2}{4\sum \alpha_\ell}}\nonumber\\
&& \left.\ \  \ \ \ \ \ \ \ \ \ \ \ \ \ \ \ \ \ \ \ \ \
e^{-\sum_{i=1}^L \alpha_i m^2 }e^{-\sum_{i=1}^L \alpha_i^{(1)} m_1^2 }e^{-\sum_{i=1}^L \alpha_i^{(2)} m_2^2 }
\right).\nonumber
\eeqa
Note that starting from the second term of the sum above, in the sum over the set of parameters $\alpha_\ell$, if the line $j_r$ belongs to that sum, then one has to make the replacement $\alpha_{j_r}\to \alpha_{j_r}+\alpha_{j_r}^{(1)}+\alpha_{j_r}^{(2)}$ in the respective $\sum \alpha_\ell$.
\end{theorem}
{\it Proof:} 
When shrinking the graph $G$, the rosette obtained is of the form of the Fig. \ref{rozeta-2P-B2}.
 The Feynman amplitude of the graph $G$ thus writes
\beqa
\label{a-plB2}
\int \prod_{i=1}^L d^D k_i e^{i(\sum_\ell k^\ell_\mu)\Theta^{\mu \nu}(\sum_k p^k_\nu)}C(k_i, m, \theta) \delta^D (\sum k,p), 
\eeqa
where the $\delta$ function expresses, as in Theorem $4.1$, the momentum conservation at each vertex. The difference with the Feynman amplitude \eqref{a-plr} of a planar regular graph comes from the oscillator factor (direct consequence of the form of the Moyal vertex). The first sum in the exponent of this new oscillator factor represents the sum on the internal momenta overarching the broken face in the rosette (in Fig. \ref{rozeta-2P-B2}, this is given by $k_1-k_3$). The second sum represents the sum over the external momenta which break the respective internal face in the rosette (in Fig. \ref{rozeta-2P-B2}, this is simply given by $p$).

Inserting  the integral representation \eqref{param} of the propagators in \eqref{a-plB2} leads to the same $2^L$ terms as in Theorem \ref{th-plr}. When performing the Gaussian integration over the internal momenta $k$, the oscillator term has to be taken into consideration, thus leading to the new type of term $e^{-\frac{\theta^2 (\sum p_k)^2}{4\sum \alpha_\ell}}$. 

Let us now prove this. Without any lose of generality, denote by $k^1$ one of the loop independent internal momentum appearing in the first sum of the exponential in \eqref{a-plB2}. We consider the first of the $2^L$ terms in the amplitude, the rest being analogous. One has to consider the following integration:
\beqa
\label{modif1}
\int d^D k^1 e^{ik^1_\mu \Theta^{\mu\nu}(\sum p^k_\nu)}\frac{1}{(k^1)^2+m^2}f(p, k^1,k).
\eeqa
The last term refers to the product of propagators of the tree lines (contracted to obtain the rosette) which contain $k_1$ (because of momentum conservation at each vertex). This product can depend, in all generality, of the external momenta $p$ as well as the internal ones, denoted here by $k$.
This product of propagators is identical to the one  of the commutative case.
Integral \eqref{modif1} further writes
\beqa
\label{modif2}
\int d^D k^1 e^{ik^1_\mu \Theta^{\mu\nu}(\sum_k p^k_\nu)}e^{-(\sum \alpha_t+\alpha_1)(k^1)^2}.
\eeqa
We have discarded the mass term as well as the linear terms in $k_1$ in the exponentials. These linear terms come from the product of propagators $f(p,k^1,k)$. 
Nevertheless, they are identical to the commutative case and do not lead to any new type of terms in the Feynman integral. The only terms to investigate are the ones of \eqref{modif2}; this integral further writes
\beqa
\label{modif3}
\int d^D k^1 
e^{-\sum \alpha_\ell (k^1_\mu - \frac{i}{2\sum \alpha_\ell}(\sum p^k_\nu \Theta^{\mu\nu})^2)}
e^{(-\sum \alpha_\ell)(\frac{i}{2\sum \alpha_\ell}(\sum p^k_\rho) \theta^{\sigma\rho})^2}
\eeqa
We have used here the simplified notation
\beqa
\sum \alpha_t + \alpha_1 = \sum \alpha_\ell.
\eeqa
The first exponential in the integral \eqref{modif3} gives the same result as in commutative field theory, while the second exponential, using the explicit form of the noncommutativity matrix \eqref{matrix}, computes straightforwardly to
\beqa
\label{new}
e^{-\frac{\theta^2 (\sum p_k)^2}{4\sum \alpha_\ell}}.
\eeqa
This completes the proof. (QED)

\medskip

The new type of term \eqref{new} can be interpreted as some kind of signature of the irregularity of the graph.
It is the only difference with respect to the parametric representation of a planar regular graph.


\subsection{Arbitrary number of broken faces}

\begin{figure}
\centerline{\epsfig{figure=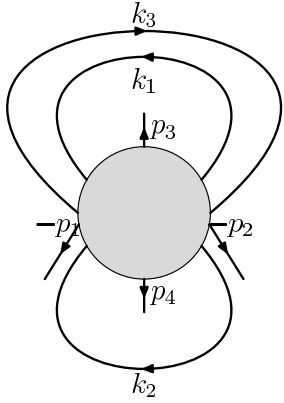,width=3.7cm} }
\caption{A four points, three broken faces rosette. Two internal faces are broken by the external momentum $p_3$ and resp. $p_4$. The first of them is overarched by the momenta $k_1-k_3$ (just like in Fig. \ref{rozeta-2P-B2}) while the second one is overarched by the momentum $-k_2$.}\label{4p}
\end{figure}

The case of an arbitrary number of broken faces is more involved. For any such internal faces $r=1, \ldots, B-1$ one has an oscillating factor in the general Feynman amplitude generalizing equation \eqref{a-plB2}.
The rosette for such a graph is of the form of Fig. \ref{4p}. In this example one has three faces. The external one is broken by $p_1+p_2$ and then one has two internal ones, broken by $p_3$ and resp. $p_4$.

The Feynman amplitude writes
\beqa
\label{a-pli}
\int \prod_{i=1}^L d^D k_i \left( \prod_{r=1}^{B-1}e^{i(\sum_\ell k^\ell_\mu)\Theta^{\mu \nu}(\sum_k p^k_\nu)}\right) C(k_i, m, \theta) \delta^D (\sum k,p). 
\eeqa
For example, in Fig. \ref{4p} one has $e^{-ip_4^\mu \Theta_{\mu\nu} k_2^\nu}$ for the first internal broken face and then $e^{-ip_3^\mu \Theta_{\mu\nu} (k_3-k_1)^\nu}$ for the second internal broken face.

One needs to perform the Gaussian integrations over the internal momenta. For the first internal broken face ($r=1$) one obtains the result of Theorem \ref{th-plB2}. For the second internal broken face ($r=2$) one obtains a factor
\beqa
e^{-\frac{\theta^2(\sum_k p_k)^2 f^{(2)}(\alpha)}{g^{(2)}(\alpha)}},
\eeqa
where $f^{(2)}(\alpha)=4\sum \alpha_\ell$ the sum of the parameters corresponding to the tree lines and to the lines overarching the first internal broken face in the rosette. Then $g^{(2)}(\alpha)$ is a positively defined polynomial of degree $2$ in the parameters $\alpha$. One has to go further iterating this process. The final result writes:

\begin{theorem}
\label{th-pli}
Let $G$ a planar regular irregular graph, with more than two broken faces.
The Feynman amplitude of $G$ writes
\beqa
\label{pli}
&&{\cal A}_G=\pi^{\frac{LD}{2}} \left( \int \prod_{i=1}^L d \alpha_i \frac{1}{[U(\alpha)]^{\frac D2}} e^{\frac{-V(\alpha, p)}{U(\alpha)}} 
\prod_{r=1}^{B-1}e^{-\frac{\theta^2 f^{(r)}(\alpha)(\sum p^{(r)}_k)^2}{g^{(r)}(\alpha)}}
e^{-\sum_{i=1}^L \alpha_i m^2 }\right.\\
&& + (-\frac{a}{\theta^2})^{L-1}\sum_{j_1=1}^L \int d\alpha_{j_1} \prod_{i\ne j_1,\, i=1}^L d\alpha_i d\alpha_i^{(1)} d\alpha_i^{(2)} \frac{1}{[U(\alpha_i+\alpha_i^{(1)}+\alpha_i^{(2)}, \alpha_{j_1})]^\frac D2} e^{-\frac{V(\alpha_i+\alpha_i^{(1)}+\alpha_i^{(2)}, \alpha_{j_1},p)}{U(\alpha_i+\alpha_i^{(1)}+\alpha_i^{(2)}, \alpha_{j_1})}} \nonumber\\
&& \ \ \ \ \ \ \ \ \ \ \ \ \ \prod_{r=1}^{B-1}e^{-\frac{\theta^2 f^{(r)}(\alpha)(\sum p^{(r)}_k)^2}{g^{(r)}(\alpha)}}
e^{-\sum_{i=1}^L \alpha_i m^2 } e^{-\sum_{i\ne j_1,\, i=1}^L \alpha_i^{(1)} m_1^2 }e^{-\sum_{i\ne j_1,\, i=1}^L \alpha_i^{(2)} m_2^2 }\nonumber\\
&& + (-\frac{a}{\theta^2})^{L-2}\sum_{j_1<j_2,\, j_1,j_2=1}^L \int d\alpha_{j_1}d\alpha_{j_2} \prod_{i\ne j_1,j_2,\, i=1}^L d\alpha_i d\alpha_i^{(1)} d\alpha_i^{(2)} \frac{1}{[U(\alpha_i+\alpha_i^{(1)}+\alpha_i^{(2)}, \alpha_{j_1}, \alpha_{j_2})]^\frac D2} \nonumber\\
&& \ \ \ \ \ \ \ \ \ \ e^{-\frac{V(\alpha_i+\alpha_i^{(1)}+\alpha_i^{(2)}, \alpha_{j_1},\alpha_{j_2}, p)}{U(\alpha_i+\alpha_i^{(1)}+\alpha_i^{(2)}, \alpha_{j_1}, \alpha_{j_2})}}
\prod_{r=1}^{B-1}e^{-\frac{\theta^2 f^{(r)}(\alpha)(\sum p^{(r)}_k)^2}{g^{(r)}(\alpha)}}
e^{-\sum_{i=1}^L \alpha_i m^2 } e^{-\sum_{i\ne j_1j_2,\, i=1}^L \alpha_i^{(1)} m_1^2 }e^{-\sum_{i\ne j_1,j_2\, i=1}^L \alpha_i^{(2)} m_2^2 }
\nonumber\\
&& + \ldots +\nonumber\\
&&  + (-\frac{a}{\theta^2})^{L}\int  \prod_{i=1}^L d\alpha_i d\alpha_i^{(1)} d\alpha_i^{(2)} \frac{1}{[U(\alpha_i+\alpha_i^{(1)}+\alpha_i^{(2)})]^\frac D2} e^{-\frac{V(\alpha_i+\alpha_i^{(1)}+\alpha_i^{(2)},p)}{U(\alpha_i+\alpha_i^{(1)}+\alpha_i^{(2)})}}
\prod_{r=1}^{B-1}e^{-\frac{\theta^2 f^{(r)}(\alpha)(\sum p^{(r)}_k)^2}{g^{(r)}(\alpha)}}\nonumber\\
&& \left.\ \  \ \ \ \ \ \ \ \ \ \ \ \ \ \ \ \ \ \ \ \ \
e^{-\sum_{i=1}^L \alpha_i m^2 }e^{-\sum_{i=1}^L \alpha_i^{(1)} m_1^2 }e^{-\sum_{i=1}^L \alpha_i^{(2)} m_2^2 }
\right).\nonumber
\eeqa
where
$f^{(r)}(\alpha)$ and resp. $g^{(r)}(\alpha)$ are positively defined polynomials in the parameters $\alpha$ of degree $r-1$ and resp. $r$.
\end{theorem}



\section{Non-planar graphs}
\label{sect-np}
\renewcommand{\theequation}{\thesection.\arabic{equation}}   
\setcounter{equation}{0}

\begin{figure}
\centerline{\epsfig{figure=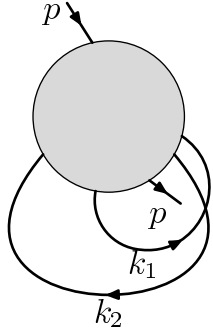,width=2.7cm} }
\caption{A non-planar rosette; the genus of the graph is $1$, as indicated by the crossing of the lines $k_1$ and $k_2$. Furthermore this ``non-planar'' face is broken by the external momentum $p$.}\label{np}
\end{figure}

In this section we  investigate the most involved case, namely the one of non-planar graphs. 
Moreover, on the rosette, one can have the face given by the non-planar lines to be one of the internal faces broken by some external leg(s) (see for example Fig. \ref{np}, where the $g=1$ value manifests by the crossing of the lines $k_1$ and $k_2$. It is this face that is further broken by the external leg). It is this more complicated case that we analyze here, with $g=1$, $B=1$
the rest of the cases going along the same lines. 
The Feynman amplitude writes
\beqa
\label{a-np}
\int \prod_{i=1}^L d^D k_i \left( e^{i((k^2-k^1)_\mu)\Theta^{\mu \nu}(\sum_k p^k_\nu)} e^{ik_1^\mu\Theta_{\mu\nu}k_2^\nu}\right) C(k_i, m, \theta) \delta^D (\sum k,p), 
\eeqa
where we have denoted by $k_1$ and $k_2$ the pair of momenta of the lines ``responsible'' for the $g=1$ non-planarity (see again Fig. \ref{np}). The sum $\sum_k p_k$ represents the sum on the external momenta of the lines which break the face of the graph (in Fig. \ref{np} this is simply given by the momentum $p$).

As before we use the integral representation \eqref{propa} of each propagator $C(k_i,m,\theta)$. We analyze here only the contribution of \eqref{a-np} given by the ``usual'' commutative propagators $C(k_i,m)$, the $2^L-1$ remaining contribution being obtained by the corresponding map of the parameters $\alpha$, as shown in the previous sections. 
One has 
\beqa
\label{a2}
\int \prod_{i=1}^L d\alpha_i d^D k_i \left( e^{i((k^2-k^1)_\mu)\Theta^{\mu \nu}(\sum_k p^k_\nu)} e^{ik_1^\mu\Theta_{\mu\nu}k_2^\nu}\right) e^{-\alpha_i (k_i^2+m^2)} \delta^D (\sum k,p). 
\eeqa
We now integrate on $k_1$. 
This leads to
\beqa
\label{11}
&&\pi^{\frac D2}\int \left (\prod_{i=1}^L d\alpha_i \right) \left(\prod_{i=2}^L d^D k_i e^{-\alpha_ik_i^2}\right) 
\frac{1}{(\alpha_1+\sum_t \alpha_t)^{\frac D2}}e^{i(k^2_\mu)\Theta^{\mu \nu}(\sum_k p^k_\nu)}
e^{\frac{-\theta^2 (\sum_k p_k^2)}{4(\alpha_1+\sum_t \alpha_t)}}
e^{\frac{-\theta^2 k_2^2}{4(\alpha_1+\sum_t \alpha_t)}}\nonumber\\
&&e^{-F(\alpha, p)}e^{-\sum_{i=1}^L \alpha_im^2},\nonumber\\
\eeqa
where $\sum_t \alpha_t$ represents the sum on the parameters $\alpha$ associated to the tree lines. (Note that neither $\alpha_1$ nor $\alpha_2$ do not belong to this sum, as can be seen again from Fig. \ref{np}).
Let us closer analyze expression \eqref{1}:
\begin{itemize}
\item
the factor $e^{\frac{-\theta^2 (\sum_k p_k^2)}{4(\alpha_1+\sum_t \alpha_t)}}$ is a noncommutativity mark containing some external momenta. It was obtained from the Gaussian integration over $e^{i((-k^1)_\mu)\Theta^{\mu \nu}(\sum_k p^k_\nu)}$; this type of term was already present at the level of planar irregular graph;
\item the function $F(\alpha, p)$ on the set of parameters $\alpha$ and the external momenta $p$ is the usual function obtained at this level of the calculus in the case of a commutative $\Phi^4$ theory;
\item the factor $e^{\frac{-\theta^2 k_2^2}{4(\alpha_1+\sum_t \alpha_t)}}$ is a noncommutativity mark containing {\it the internal momentum} $k_2$. This is a new type of factor which is obtained from the Gaussian integration over  $e^{ik_1^\mu\Theta_{\mu\nu}k_2^\nu}$; this type of factor is a {\it mark of the non-planarity} of the graph.
\end{itemize}

We now show that when integrating over $k_2$, this new type of term will lead to an {\it explicit dependence on $\theta$ of the first polynomial}, which was not the case for planar (regular or irregular) graphs. Indeed, \eqref{11} leads to
\beqa
\label{111}
&&\pi^{D}\int \left (\prod_{i=1}^L d\alpha_i \right) \left(\prod_{i=3}^L d^D k_i e^{-\alpha_ik_i^2}\right) 
\frac{1}{(\alpha_1+\sum_t \alpha_t)^{\frac D2}(\frac{f_2(\alpha)}{\alpha_1+\sum_t \alpha_t}+\frac{\theta^2}{4(\alpha_1+\sum_t \alpha_t)})^{\frac D2}}\nonumber\\
&&e^{-G(\alpha, p, k_3, \ldots, k_L,\theta)}e^{-\sum_{i=1}^L \alpha_im^2},
\eeqa
where $f_2(\alpha)$ is some function of degree $2$ on the parameters $\alpha_t$, $\alpha_1$ and $\alpha_2$ (which would have been obtained at this level of the calculations also in the commutative $\Phi^4$ case) and $G(\alpha, p,\theta)$ is some function on the set of parameters $\alpha$, the external momenta $p$, the noncommutativity $\theta$ and finally the remaining internal momenta $k_3,\ldots, k_L$.

The rest of the Gaussian integrations proceeds as in the commutative $\Phi^4$ case (since we have already integrated over both noncommutativity marks $e^{i((k^2-k^1)_\mu)\Theta^{\mu \nu}(\sum_k p^k_\nu)}$ and $ e^{ik_1^\mu\Theta_{\mu\nu}k_2^\nu}$, see the initial equation \eqref{a-np}). Finally, one obtains
\beqa
\label{a-np-final}
&&\pi^{\frac{LD}{2}}\int \left (\prod_{i=1}^L d\alpha_i \right)  
\frac{1}{(\alpha_1+\sum_t \alpha_t)^{\frac D2}(\frac{f_2(\alpha)}{\alpha_1+\sum_t \alpha_t}+\frac{\theta^2}{4(\alpha_1+\sum_t \alpha_t)})^{\frac D2}(\frac{f_3(\alpha)}{f_2(\alpha)})^{\frac D2}\ldots (\frac{f_L (\alpha)}{f_{L-1}(\alpha)})^{\frac D2}}\nonumber\\
&&e^{-G(\alpha, p,\theta)}e^{-\sum_{i=1}^L \alpha_im^2},
\eeqa
where $f_j(\alpha)$ is a function of degree $j$ in the parameters $\alpha$, function which would have been obtained also in the case of the Gaussian integrations in the commutative $\Phi^4$ model. Note that $f_L(\alpha)= U(\alpha)$ as defined in the commutative case in Definition \ref{polinoame}.  

\medskip

Before ending this section, let us stress that the main difference with the parametric representation of the planar (regular or irregular) graphs is  the presence of a noncommutativity  trace, the $\theta^2$ terms, already at the level of the first polynomial.

\section{Comments; structure of divergences and power counting dependence on the genus}
\label{sect-modif}
\renewcommand{\theequation}{\thesection.\arabic{equation}}   
\setcounter{equation}{0}

In this section we make some further comments with respect to the parametric representation we have implemented in this paper. We also analyze the structure of divergences in the parametric space. Furthermore we explicitly calculate the superficial degree of divergence as a function of the graph genus. This is an improvement with respect to the result obtained in \cite{noi}.

\medskip



Let us investigate the structure of divergences of the first of the $2^L$ terms of the amplitude (the one corresponding to the use of only ``usual'', commutative, propagators).

For the planar regular sector, the superficial degree of divergence is obtained analogously to the case of a commutative theory (because the oscillating Moyal factors do not interfere in the Feynman integrals, as already explained in section $4$).

For the planar irregular sector, the situation is identical with the exception of a supplementary term of type \eqref{new}. This term improves the UV convergence of the integrand. In \cite{limita-GMRT} it was shown, using the form of the Feynman integral in the parametric space, that such integrals are finite.

Let us investigate in greater detail the behavior of the Feynman integrals for the case of non-planar graphs. One notices that the $\theta^2$ terms affect the homogeneity in the $\alpha$ parameters. Indeed, all the terms except the $\theta^2$ terms in the polynomial are of total degree in the $\alpha$ parameters:
\beqa
L-(n-1).
\eeqa
Each time one has a pair of genus lines on the rosette ({\it i. e.} the graph is non-planar), we have seen that we have a $\theta^2$ term in the first polynomial. One thus has a term with a  minimal total degree in the $\alpha$ parameters 
\beqa
\label{min}
L-(n-1)-2g.
\eeqa

To obtain the power counting one makes the usual rescaling
\beqa
\alpha \to \rho \alpha.
\eeqa
In the UV regime ($\alpha\to 0$), the dominant term is the term with maximal degree of $\theta$ (and hence minimal degree \eqref{min} in the $\alpha$ parameters). 
The superficial degree of divergence (given by the rescaling parameter $\rho$) is thus
\beqa
\label{aproape}
\omega = L- 2(L-(n-1)-2g)= -L+2n-2+4g.
\eeqa
We now use the general relation
\beqa
\label{aj}
L=2n-\frac N2,
\eeqa
where $N$ is the number of external legs. Inserting \eqref{aj} in \eqref{aproape} leads to the result:
\beqa
\label{im}
\omega=\frac 12 (N-4)+4g.
\eeqa

\medskip

Let us argue that the rest of $2^L-1$ terms of the Feynman amplitude (which imply the use of at least one noncommutative correction for some propagator)  are convergent in the UV. When one replaces the ``usual'' commutative propagator by its noncommutative correction, $2$ supplementary integrations on some additional $\alpha$ parameters are introduced. Nevertheless, the homogeneity and the total degree in the $\alpha$ parameters of the polynomials are not changed. This improvement with respect to the case of the first term of the amplitude (treated above) leads to the fact that these $2^L-1$ corrections are irrelevant in the UV. Note that this result was already obtained by $2$ different methods in {\it momentum space} in \cite{beta-GMRT}.

\medskip

We end this section by emphasizing that the explicit genus dependence of the power counting \eqref{im} is, as already stated above, an improvement with respect to the result of \cite{noi}, where only a limit on $\omega$ was proved for the case of non-planar graphs.

\section{Examples}
\label{sect-exemple}
\renewcommand{\theequation}{\thesection.\arabic{equation}}   
\setcounter{equation}{0}

\begin{figure}
\centerline{\epsfig{figure=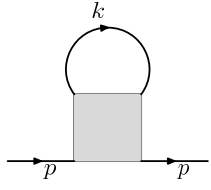,width=2.7cm} }
\caption{A planar tadpole graph.}\label{tad}
\end{figure}

In this section we explicitly calculate the parametric representation of some particular Feynman graphs. 
We represent the ribbon graphs as ``usual'' graphs, but with a non-local vertices.
We start by the simplest possible graph, the planar regular tadpole (see Fig. \ref{tad}).
One has
\beqa
\int d^4 k C(k,m,\theta)
\eeqa
Inserting  the integral representation \eqref{param} of the propagator $C(k,m,\theta)$, one has
\beqa
\label{1}
\int d\alpha d^Dk e^{-\alpha (k^2+m^2)}
- \frac{a}{\theta^2} \int d\alpha d\alpha^{(1)} d\alpha^{(2)} d^Dk e^{-(\alpha+\alpha^{(1)}+\alpha^{(2)})k^2} e^{-\alpha m^2} e^{-\alpha^{(1)} m_1^2}e^{-\alpha^{(2)} m_2^2}.
\eeqa
Performing  the Gaussian integral on $k$ in both terms above leads to
\beqa
\pi^{\frac D2}
\int d\alpha \frac{1}{\alpha^{\frac D2}} e^{-\alpha m^2}
- \pi^{\frac D2}\frac{a}{\theta^2} \int d\alpha d\alpha^{(1)} d\alpha^{(2)} 
\frac{1}{(\alpha+\alpha^{(1)}+\alpha^{(2)})^{\frac D2}}  
e^{-\alpha m^2} e^{-\alpha^{(1)} m_1^2}e^{-\alpha^{(2)} m_2^2}.
\eeqa
This allows to identify the expected results
\beqa
U(\alpha)=\alpha,\ \ V=0.
\eeqa

\medskip

\begin{figure}
\centerline{\epsfig{figure=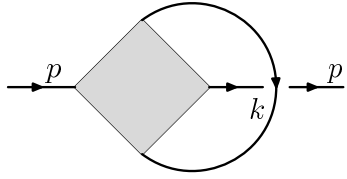,width=4.5cm} }
\caption{The planar irregular tadpole graph.}\label{np-tad}
\end{figure}

Let us go further to the planar irregular version of this graph (see Fig. \ref{np-tad}). 
One has
\beqa
\int d^D k e^{ik_\mu \Theta^{\mu\nu} p_\nu} C(k, m, \theta).
\eeqa
Inserting the integral representation \eqref{param} of the propagator $C(k,m,\theta)$, one has
\beqa
&&\int d\alpha d^Dk e^{-\alpha (k^2+m^2)} e^{ik_\mu \Theta^{\mu\nu} p_\nu}\nonumber\\
&&- \frac{a}{\theta^2} \int d\alpha d\alpha^{(1)} d\alpha^{(2)} d^Dk e^{-(\alpha+\alpha^{(1)}+\alpha^{(2)})k^2} e^{ik_\mu \Theta^{\mu\nu} p_\nu} e^{-\alpha m^2} e^{-\alpha^{(1)} m_1^2}e^{-\alpha^{(2)} m_2^2}.\nonumber
\eeqa
The Gaussian integral has to take into consideration also the oscillating factor $e^{ik_\mu \Theta^{\mu\nu} p_\nu}$. One has
\beqa
&&\pi^{\frac D2}\int d\alpha \frac{1}{\alpha^{\frac D2}} e^{-\frac{\theta^2p^2}{4\alpha}}e^{-\alpha m^2}\\
&&- \pi^{\frac D2}\frac{a}{\theta^2} \int d\alpha d\alpha^{(1)} d\alpha^{(2)} 
\frac{1}{(\alpha+\alpha^{(1)}+\alpha^{(2)})^{\frac D2}}   e^{-\frac{\theta^2p^2}{4(\alpha+\alpha^{(1)}+\alpha^{(2)})}}
e^{-\alpha m^2} e^{-\alpha^{(1)} m_1^2}e^{-\alpha^{(2)} m_2^2}.\nonumber
\eeqa
This expression provides the parametric representation.

\medskip

\begin{figure}
\centerline{\epsfig{figure=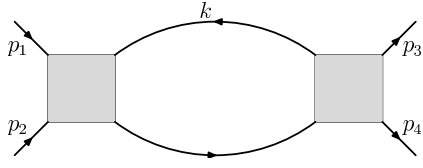,width=5.5cm} }
\caption{The bubble graph.}\label{fig-bubble}
\end{figure}

Let us investigate a more elaborated graph, namely the bubble graph of Fig. \ref{fig-bubble}. 
One has to deal with the integral
\beqa
\label{bubble}
\int d^D k C(k, m, \theta) C(k+P, m, \theta),
\eeqa
where $P=p_1+p_2$. Inserting the integral representation \eqref{param} leads to
\beqa
\int d^D k 
\left( \int d\alpha_1 e^{-\alpha_1 (k^2+m^2)}
- \frac{a}{\theta^2} \int d\alpha_1 d\alpha_1^{(1)} d\alpha_1^{(2)} e^{-(\alpha_1+\alpha_1^{(1)}+\alpha_1^{2)})k^2} e^{-\alpha_1 m^2} e^{-\alpha_1^{(1)} m_1^2}e^{-\alpha_1^{(2)} m_2^2}\right.\nonumber\\
\left. \int d\alpha_2 e^{-\alpha_1 ((k+P)^2+m^2)}
- \frac{a}{\theta^2} \int d\alpha_2 d\alpha_2^{(1)} d\alpha_2^{(2)} e^{-(\alpha_2+\alpha_2^{(1)}+\alpha_2^{2)})(k+P)^2} e^{-\alpha_2 m^2} e^{-\alpha_2^{(1)} m_1^2}e^{-\alpha_2^{(2)} m_2^2}\right).\nonumber\\
\eeqa
Performing the Gaussian integration finally leads to
\beqa
&&\pi^{\frac D2}\left(\int d\alpha_1 d\alpha_2 \frac{1}{(\alpha_1+\alpha_2)^{\frac D2}} e^{-\frac{\alpha_1 \alpha_2P^2}{\alpha_1+\alpha_2}}e^{-(\alpha_1+\alpha_2)m^2}\right.\\
&&+ \int d\alpha_1 d\alpha_2 d\alpha_2^{(1)}d\alpha_2^{(2)} \frac{1}{(\alpha_1+\alpha_2+\alpha_2^{(1)}+\alpha_2^{(2)})^{\frac D2}} e^{-\frac{\alpha_1 (\alpha_2+\alpha_2^{(1)}+\alpha_2^{(2)}P^2}{\alpha_1+\alpha_2+\alpha_2^{(1)}+\alpha_2^{(2)}}}e^{-(\alpha_1+\alpha_2)m^2+\alpha_2^{(1)}m_1^2+\alpha_2^{(2)}m^2_2}\nonumber\\
&&+ \int d\alpha_2 d\alpha_1 d\alpha_1^{(1)}d\alpha_1^{(2)} \frac{1}{(\alpha_2+\alpha_1+\alpha_1^{(1)}+\alpha_1^{(2)})^{\frac D2}} e^{-\frac{\alpha_2 (\alpha_1+\alpha_1^{(1)}+\alpha_1^{(2)}P^2}{\alpha_2+\alpha_1+\alpha_1^{(1)}+\alpha_1^{(2)}}}e^{-(\alpha_1+\alpha_2)m^2+\alpha_1^{(1)}m_1^2+\alpha_1^{(2)}m^2_2}\nonumber\\
&&\left.+ \int d\alpha_1 d\alpha_1^{(1)}d\alpha_1^{(2)}d\alpha_2 d\alpha_2^{(1)}d\alpha_2^{(2)} \frac{1}{(\alpha_1+\alpha_1^{(1)}+\alpha_1^{(2)}+\alpha_2+\alpha_2^{(1)}+\alpha_2^{(2)})^{\frac D2}} e^{-\frac{(\alpha_1+\alpha_1^{(1)}+\alpha_1^{(2)}) (\alpha_2+\alpha_2^{(1)}+\alpha_2^{(2)}P^2}{\alpha_1+\alpha_1^{(1)}+\alpha_1^{(2)}+\alpha_2+\alpha_2^{(1)}+\alpha_2^{(2)}}}\right.\nonumber\\
&& \ \ \ \ \ \ \ \ \ \ \left. e^{-(\alpha_1+\alpha_2)m^2+(\alpha_1^{(1)}+(\alpha_1^{(1)}+\alpha_2^{(1)})m_1^2+(\alpha_1^{(2)}+\alpha_2^{(2)}m^2_2}\right).\nonumber
\eeqa
This allows to identify the polynomials
\beqa
U(\alpha)=\alpha_1+\alpha_2, \ \ V(\alpha, p_1,p_2)= \alpha_1 \alpha_2 P^2,
\eeqa
where $P$ is the total incoming (or outgoing momentum).

\begin{figure}
\centerline{\epsfig{figure=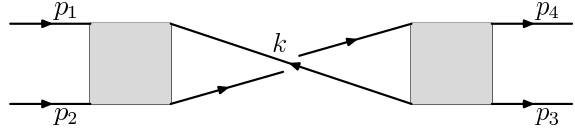,width=7.5cm} }
\caption{A planar ($g=0$) irregular ($B=2$) example of Feynman graph.}\label{ex4PB2}
\end{figure}

When switching the ending points of the graph Fig. \ref{fig-bubble}, one obtains 
the planar irregular ($B=2$) graph of Fig. \ref{ex4PB2}. Its amplitude contains an extra oscillator factor 
\beqa
\label{bubbleB2}
\int d^D k e^{ik_\mu \Theta^{\mu \nu}P_\nu}C(k, m, \theta) C(k+P, m, \theta).
\eeqa
The calculation proceeds along the same lines leading to
\beqa
&&\pi^{\frac D2} \left(\int d\alpha_1 d\alpha_2 \frac{1}{(\alpha_1+\alpha_2)^{\frac D2}} e^{-\frac{\alpha_1 \alpha_2P^2}{\alpha_1+\alpha_2}}
e^{-\frac{\theta^2 P^2}{4(\alpha_1+\alpha_2)}}
e^{-(\alpha_1+\alpha_2)m^2}\right.\\
&&+ \int d\alpha_1 d\alpha_2 d\alpha_2^{(1)}d\alpha_2^{(2)} \frac{1}{(\alpha_1+\alpha_2+\alpha_2^{(1)}+\alpha_2^{(2)})^{\frac D2}} e^{-\frac{\alpha_1 (\alpha_2+\alpha_2^{(1)}+\alpha_2^{(2)})P^2}{\alpha_1+\alpha_2+\alpha_2^{(1)}+\alpha_2^{(2)}}}
e^{-\frac{\theta^2 P^2}{4(\alpha_1+\alpha_2+\alpha_2^{(1)}+\alpha_2^{(2)})}}\nonumber\\
&& \ \ \ \ \ \ \ \ \ \ \ \ \ \ \ \ \ \ \ \ \ \ \
e^{-(\alpha_1+\alpha_2)m^2+\alpha_2^{(1)}m_1^2+\alpha_2^{(2)}m^2_2}\nonumber\\
&&+ \int d\alpha_2 d\alpha_1 d\alpha_1^{(1)}d\alpha_1^{(2)} \frac{1}{(\alpha_2+\alpha_1+\alpha_1^{(1)}+\alpha_1^{(2)})^{\frac D2}} e^{-\frac{\alpha_2 (\alpha_1+\alpha_1^{(1)}+\alpha_1^{(2)})P^2}{\alpha_2+\alpha_1+\alpha_1^{(1)}+\alpha_1^{(2)}}}
e^{-\frac{\theta^2 P^2}{4(\alpha_2+\alpha_1+\alpha_1^{(1)}+\alpha_1^{(2)})}}\nonumber\\
&& \ \ \ \ \ \ \ \ \ \ \ \ \ \ \ \ \ \ \ \ \ \ \
e^{-(\alpha_1+\alpha_2)m^2+\alpha_1^{(1)}m_1^2+\alpha_1^{(2)}m^2_2}\nonumber\\
&&\left.+ \int d\alpha_1 d\alpha_1^{(1)}d\alpha_1^{(2)}d\alpha_2 d\alpha_2^{(1)}d\alpha_2^{(2)} \frac{1}{(\alpha_1+\alpha_1^{(1)}+\alpha_1^{(2)}+\alpha_2+\alpha_2^{(1)}+\alpha_2^{(2)})^{\frac D2}} e^{-\frac{(\alpha_1+\alpha_1^{(1)}+\alpha_1^{(2)}) (\alpha_2+\alpha_2^{(1)}+\alpha_2^{(2)})P^2}{\alpha_1+\alpha_1^{(1)}+\alpha_1^{(2)}+\alpha_2+\alpha_2^{(1)}+\alpha_2^{(2)}}}\right.\nonumber\\
&& \ \ \ \ \ \ \ \ \ \ \left. 
e^{-\frac{\theta^2 P^2}{4(\alpha_1+\alpha_1^{(1)}+\alpha_1^{(2)}+\alpha_2+\alpha_2^{(1)}+\alpha_2^{(2)})}}
e^{-(\alpha_1+\alpha_2)m^2+(\alpha_1^{(1)}+(\alpha_1^{(1)}+\alpha_2^{(1)})m_1^2+(\alpha_1^{(2)}+\alpha_2^{(2)}m^2_2}\right).\nonumber
\eeqa
This completes the parametric representation.

\medskip

\begin{figure}
\centerline{\epsfig{figure=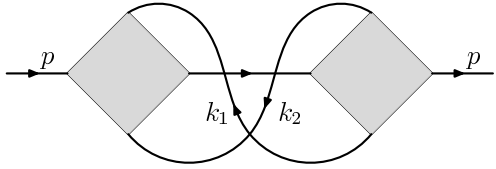,width=6.5cm} }
\caption{An example of a non-planar ($g=1$) graph.}\label{fig-np}
\end{figure}

Let us end our series of examples by the non-planar graph of Fig. \ref{fig-np}. This graph has $g=1$.

Its Feynman amplitude writes
\beqa
\int d^D k_1 d^D k_2 e^{-i(p_\mu \Theta^{\mu \nu} (k_2-k_1)_\nu + k_1^\mu \Theta_{\mu \nu}k_2^\nu)}C(k_1, m, \theta) C(k_2, m,\theta) C(p+k_1+k_2,m,\theta).
\eeqa
As before one inserts the integral representation \eqref{param} for the propagators. We give here only the contribution obtained from $C(k_1, m) C(k_2, m) C(p+k_1+k_2,m)$, the rest of the $7$ terms being obtained from the respective replacements of the set of parameters $\alpha$, as described throughout this paper. After performing first the Gaussian integration on $k_1$ and then on $k_2$ one obtains for the first polynomial the expression
\beqa
\alpha_1\alpha_2+\alpha_1\alpha_3+\alpha_2\alpha_3+\frac 14 \theta^2.
\eeqa
The $\theta^2$ factor is a direct manifestation of the non-planarity of the graph.

Let us end remark that, in the limit $\theta\to 0$, one obtains, as expected, the polynomials of the commutative graph (since in this case, the non-planarity plays no r\^ole).

\bigskip


We have thus established the parametric representation for all type of Feynman graphs, planar (regular or irregular) or non-planar. As shown above, this allows to obtain the power counting of our model. The dependence of the superficial degree of divergence $\omega$ on the graph genus has been explicitly found. This is in improvement of the result obtained in \cite{noi}, where only a bound on $\omega$ has been proved (for the case of non-planar graphs).





Furthermore, this parametric representation can be the starting point for the definition of the Mellin representation of this noncommutative model.
We finally remark that, as in the commutative case, the parametric representation we have implemented in this paper has the space dimension $D$ as a simple parameter in the formulas. Thus, this is a natural way to implement the dimensional regularization, as was done for the Grosse-Wulkenhaar model in \cite{dimreg}.

\bigskip

{\bf Acknowledgment:} 
The author acknowledges the CNCSIS grant ``Idei'' 454/2009, 
ID-44 for partial support.

\end{document}